\documentclass[twocolumn]{jpsj2} 

\title{
Possible Three-Dimensional Nodes 
in the $s{\pm}$ Superconducting Gap of 
BaFe$_2$(As$_{1-x}$P$_x$)$_2$}
\author{
Katsuhiro Suzuki$^{1,2}$, Hidetomo Usui$^1$, and Kazuhiko Kuroki$^{1,2}$ }
\inst{
$^1$ Department of Applied Physics and Chemistry, 
The University of Electro-Communications, Chofu, Tokyo 182-8585, Japan\ \\
$^2$ JST, TRIP, Chofu, Tokyo 182-8585, Japan}
\abst{
We theoretically examine the superconducting state of 
BaFe$_2$(As$_{1-x}$P$_x$)$_2$, an isovalent doping 122-iron-pnictide 
superconductor. 
We construct a three-dimensional ten-orbital model by first-principles 
band calculation, and investigate the superconducting gap within the 
spin-fluctuation-mediated pairing mechanism.  
The gap is basically $s\pm$, where the gap changes its sign between 
electron and hole Fermi surfaces, but 
three-dimensional nodal structures appear in the 
largely warped 
hole Fermi surface having a strong $Z^2/XZ/YZ$ orbital character. 
The present result, together with our previous study of the 1111 systems, 
explains the strong material dependence of the superconducting gap 
in iron pnictides.}
\kword{122 iron pnictides, superconducting gap, spin fluctuations, 
first-principles band calculation}

\begin{document}
\maketitle

It is now becoming clearer that 
the superconducting gap of iron pnictides\cite{review} 
has nonuniversal forms.
A number of experiments have suggested the presence of a fully open 
gap\cite{Matsuda,Ding},
but for LaFePO in particular, 
experiments have suggested the presence of line nodes in the 
gap\cite{Fletcher,Hicks,Yamashita}. 
In a previous study, motivated by an experimental observation by 
Lee {\it et al.}\cite{LeeStruct}, 
we have mainly focused on the  1111 systems, and 
pointed out that the band structure 
around the wave vector $(\pi,\pi)$ in the 
unfolded Brillouin zone, and thus superconductivity, is 
sensitive to the pnictogen height measured from the iron plane\cite{KKheight}. 
Using a five-orbital model\cite{KK1st}, we have shown that 
when pnictogen is at high positions, the 
$X^2-Y^2$ band around $(\pi,\pi)$ gives a hole Fermi surface, 
and in this case, the spin fluctuations arising from the interaction 
between electron and hole Fermi surfaces give rise to a fully gapped 
$s\pm$-wave pairing\cite{Mazin,KK1st,Ikeda0,Nomura}.
On the other hand, as the pnictogen is lowered, 
the $Z^2$ band around $(\pi,\pi)$ rises up above the Fermi level, 
while the $X^2-Y^2$ band sinks below.
This results in the presence of line nodes in the superconducting 
gap either in the nodal $s\pm$-wave form, where the gap nodes are on the 
electron Fermi surfaces, or in the $d$-wave form where the nodes are on the 
hole Fermi surfaces\cite{Graser,KK1st}. 
Fluctuation exchange (FLEX)\cite{Ikeda1} and 
functional renormalization group\cite{DHLee,Thomale} 
studies also found a similar tendency.
The presence of line nodes in the superconducting gap results in a 
lower $T_c$, and this explains the experimental observation of 
line nodes in low-$T_c$ LaFePO\cite{Fletcher,Hicks,Yamashita}.

However, if we focus on the 122 materials, 
there are now several experiments that do not fit into this view.
In the isovalent doping system BaFe$_2$(As$_{1-x}$P$_x$)$_2$
\cite{Ren,Kasahara}, 
a number of experiments suggested the 
presence of line nodes in the 
superconducting gap\cite{Hashimoto,Nakai}, but $T_c$ 
is relatively high (maximum $T_c$ of about 30 K)\cite{Kasahara}, 
in contrast to LaFePO.
Moreover, the $X^2-Y^2$ hole Fermi surface is found to be 
present in this material theoretically (as seen below), and 
this also seems to be supported by results of an angle resolved 
photoemission (ARPES) experiment\cite{Shimojima}.
Another interesting observation in the 122 materials 
is the possible presence of 
nodes in the superconducting gap of KFe$_2$As$_2$
\cite{Fukazawa,Dong,Furukawa,Zhang,Matsuda2}. In this material, the electron 
Fermi surface is barely present\cite{Sato}, so it is likely that the 
gap nodes are on the hole Fermi surface. Here also, the $X^2-Y^2$ hole Fermi 
surface should be present because there is a large number of holes.
These experimental observations for the 122 materials do not seem to fit 
into the view that nodal pairing occurs 
when the $X^2-Y^2$ hole Fermi surface disappears 
by reducing the pnictogen height. 
As for other origins of the presence of line nodes in the gap, 
the Coulomb avoidance\cite{Chubukov}, as well as the competition of spin and 
orbital fluctuations\cite{Kontani,Ohno}, has been considered.

In the present study, we consider another possible origin of the 
gap nodes, which is peculiar to the 122 materials. We construct 
{\it a three-dimensional ten-orbital model} of BaFe$_2$(As$_{1-x}$P$_x$)$_2$ 
by first-principles calculation using 
maximally localized Wannier orbitals\cite{MaxLoc}, 
and apply random phase approximation (RPA)
to obtain the spin susceptibility and superconducting gap function. 
For the 122 materials, the Brillouin zone unfolding procedure\cite{KK1st} 
that adopts the reduced unit cell (with one iron) cannot be performed 
strictly\cite{Mazin122} there fore, 
by adopting the ten-orbital model that uses the original unit cell 
(with two irons) of the body-centered tetragonal lattice structure,
we fully take into account the peculiar features of the 122 band structure 
not present in the 1111 materials.
The superconducting gap is basically $s_\pm$, 
where the gap changes its sign between 
electron and hole Fermi surfaces, but 
when the hole Fermi surface around the Z point 
having a strong $3Z^2-R^2$ ($Z^2$) orbital character becomes large by 
isovalent doping, 
the superconducting gap on that Fermi surface exhibits three-dimensional 
nodal structures. This kind of node is peculiar to 
the 122 systems, as also found in a five-orbital study of 
BaFe$_2$As$_2$ (but with different orbital characters of 
the Fermi surface)\cite{Graser122,Hirschfeld}. 
Since the $Z^2$ orbital does not play an important role in 
spin-fluctuation-mediated superconductivity, this explains why
$T_c$ in the isovalent doping system is relatively high 
despite the presence of nodes in the superconducting gap.

\begin{figure}[t]
\begin{center}
\includegraphics[width=8cm,clip]{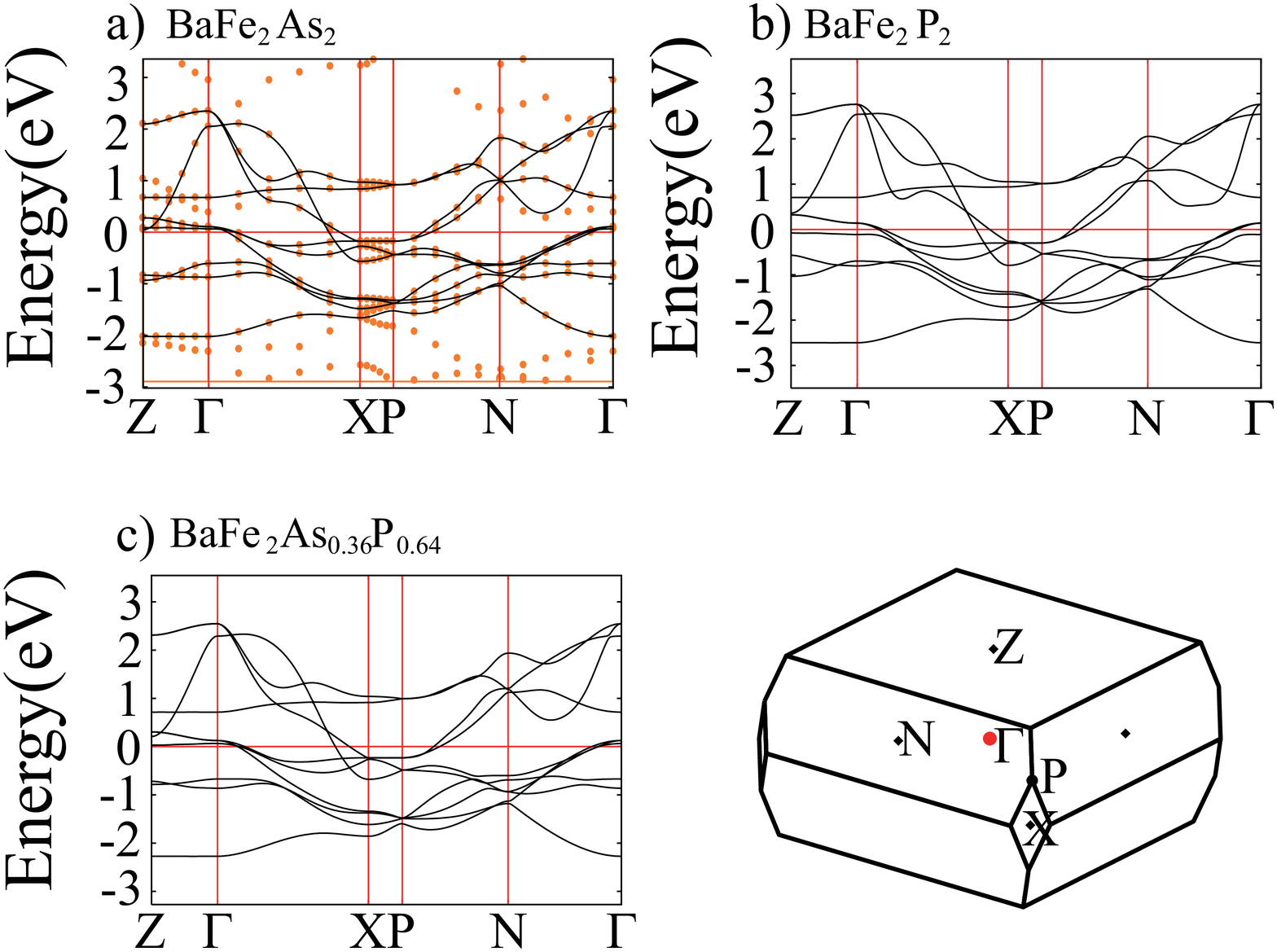}
\end{center}
\caption{(Color online) Band structures of the ten-orbital model of 
(a) BaFe$_2$As$_2$, (b) BaFe$_2$P$_2$, and 
(c) BaFe$_2$As$_{0.36}$P$_{0.64}$. In (a), the original 
first-principles band calculation is shown by dots. The 
Brillouin zone is shown in the inset.
}
\label{fig1}
\end{figure}

In order to construct a realistic model, 
we first perform first-principles band calculation 
using the Quantum Espresso package\cite{pwscf} adopting the 
experimentally determined lattice structure\cite{struct}, 
and then obtain a ten-orbital model 
using maximally localized Wannier orbitals\cite{MaxLoc}.
There are ten bands mainly originating from the five 3d orbitals because
there are two iron atoms per unit cell in the body-centered tetragonal 
lattice structure.
We show in the upper panels of Fig. \ref{fig1}
the band structure of the ten-orbital models of 
BaFe$_2$As$_2$ and BaFe$_2$P$_2$.  
To obtain a model of BaFe$_2$(As$_{1-x}$P$_x$)$_2$, 
we first obtain a ten-orbital tight-binding model of 
``hypothetical'' BaFe$_2$As$_2$ and BaFe$_2$P$_2$
having the experimentally determined lattice structure 
of BaFe$_2$(As$_{1-x}$P$_x$)$_2$ at each $x$. 
We assume that making a linear 
combination of the two sets of tight-binding 
parameters (hopping integrals and on-site energies) and mixing them 
with a ratio of $1-x:x$ will give a good approximation of the band 
structure of BaFe$_2$(As$_{1-x}$P$_x$)$_2$.
The band structure of this model with $x=0.64$ is 
shown in the lower panel of Fig. \ref{fig1}.

\begin{figure}[t]
\begin{center}
\includegraphics[width=8cm,clip]{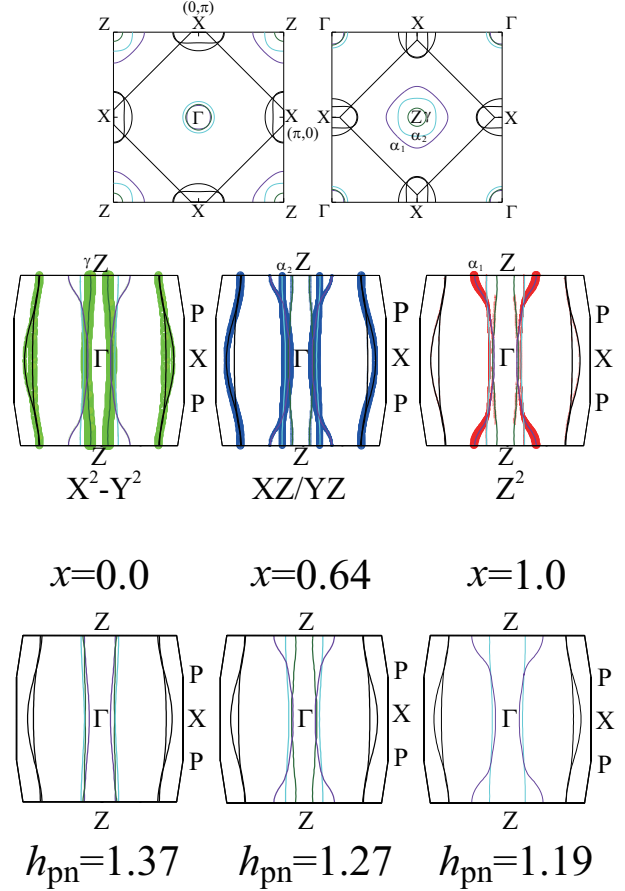}
\end{center}
\caption{(Color online) Upper panels: Fermi surfaces of the ten-orbital model of 
BaFe$_2$(As$_{0.36}$P$_{0.64}$)$_2$. Horizontal (vertical) 
cuts are shown in the top (middle) panels. 
The vertical cuts are presented along with the 
strength of the orbital character. The X point 
corresponds to the wavevector $(\pi,0)$  in the 
unfolded Brillouin zone.
Bottom panels: evolution of the Fermi surface upon increasing the 
phosphoruss concentration $x$. The pnictogen height $h_{\rm Pn}$ 
is also presented.}
\label{fig2}
\end{figure}

In Fig. \ref{fig2}, we show the Fermi surfaces of 
BaFe$_2$(As$_{0.36}$P$_{0.64}$)$_2$ for 
each orbital character, where the thickness represents the strength of the 
character.
The 122 materials share common features with the 1111 materials in that 
they have three hole (around $\Gamma$-Z) 
and two electron Fermi (around X-P) surfaces. There are,
however, some differences. 
One is that the portion of the 
Fermi surface around the Z point 
having a strong $Z^2$ character  is continuously 
connected to the $XZ/YZ$ portion of the Fermi surface around the
$\Gamma$ point in 122, while in 1111 the $Z^2$ Fermi surface, when present, 
is an isolated three-dimensional pocket. 
We will call this hole Fermi surface with a mixed $Z^2$ and $XZ/YZ$ 
orbital character $\alpha_1$. 
Another difference of 122 from 1111 is that the hole Fermi surface 
having a strong $X^2-Y^2$ orbital character coexists with the $Z^2$ 
Fermi surface; in 1111, either the three-dimensional $Z^2$  or the 
cylindrical $X^2-Y^2$ ($\gamma$) Fermi surfaces exist 
depending on the pnictogen height\cite{KKheight,Singh,Lebegue,GeorgesArita}.
The hole Fermi surface with an $X^2-Y^2$ 
character will be called $\gamma$, as in our study of 1111\cite{KKheight}.
There is another 
hole Fermi surface having an $XZ/YZ$ character, which we will call 
$\alpha_2$. 

Replacing As by P does not alter the band filling (isovalent doping), and the 
main effect is to reduce the pnictogen height measured from the 
iron planes. In the bottom panels of Fig. \ref{fig2}, 
we show the vertical cut of the 
Fermi surface for various P contents. As in 1111, 
lowering the height leads to a higher $Z^2$ orbital level, 
resulting in a larger 
$\alpha_1$ Fermi surface around the Z point\cite{Kasahara}. 
Such a strong warping of the Fermi surface has been directly observed in a
recent ARPES experiment\cite{Yoshida}.
At the same time, the $X^2-Y^2$ level is lowered, so that the 
$\gamma$  Fermi surface shrinks, 
but it exists up to a P content of about $x\sim 0.7$ in the present 
calculation.

We now move on to the RPA calculation. We mainly concentrate on 
BaFe$_2$(As$_{0.36}$P$_{0.64}$)$_2$ here, while we also comment on the 
calculation results for other $x$ values.
As for the electron-electron interaction, we consider the intraorbital
$U$, the interorbital $U'$, Hund's coupling $J$, and the pair hopping
interaction $J'$. We consider the orbital-dependent interactions 
obtained in ref.\cite{Miyake} . 
We apply RPA in this model\cite{KK1st,KKheight}, and obtain the 
spin and charge susceptibility matrices. From these, we obtain the 
pairing interaction, which are plugged into the 
linearized Eliashberg equation. 
In the RPA calculation
(where the self-energy correction is neglected), 
realistic values of the interaction results in very large spin 
fluctuations, so we multiply all the electron-electron 
interaction by a factor $f$\cite{KKheight}. In this way, we keep the 
relative strength of the interactions between different orbitals to be 
the same as those obtained from the first-principles calculation.
In the following, we will present the eigenfunction of the Eliashberg
equation in the band representation, and call them 
the ``(superconducting) gap''.  
In the actual calculation results shown below, 
we take $16\times 16\times 16$ $k$-point meshes, 
128 Matsubara frequencies, $T=0.07$ eV, and the interaction 
reducing ratio $f=0.55$.

\begin{figure}
\begin{center}
\includegraphics[width=5cm,clip]{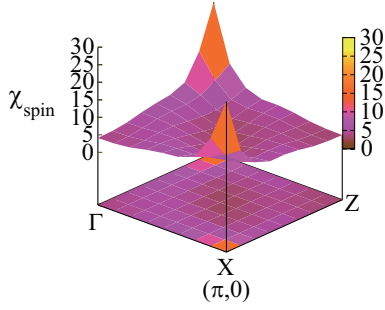}
\end{center}
\caption{(Color online) Largest eigenvalue of the spin susceptibility matrix for the 
ten-orbital model of BaFe$_2$(As$_{0.36}$P$_{0.64}$)$_2$.
The X point corresponds to the $(\pi,0)$ point in the unfolded 
Brillouin zone.}
\label{fig3}
\end{figure}

In Fig. \ref{fig3}, we show a plot of the eigenvalue of the 
spin susceptibility matrix. It has a peak at the X point  
(corresponding to the wave vector $(\pi,0)$ in the unfolded 
Brillouin zone, although the unfolding cannot be strictly performed), 
which originates 
from the interaction between electron and hole Fermi surfaces.
As shown in our earlier study\cite{KKheight}, 
the main interaction that induces the spin fluctuation comes from the 
$X^2-Y^2$ and $XZ/YZ$ portions of the Fermi surface, namely, the 
intraorbital interaction within these orbitals. 
Performing a similar calculation for other values of $x$, we find that 
the spin susceptibility is suppressed monotonically upon 
increasing the phosphoruss content, as expected from the fact that the 
$\gamma$ Fermi surface shrinks\cite{KKheight}.

\begin{figure}[t]
\begin{center}
\includegraphics[width=7.7cm,clip]{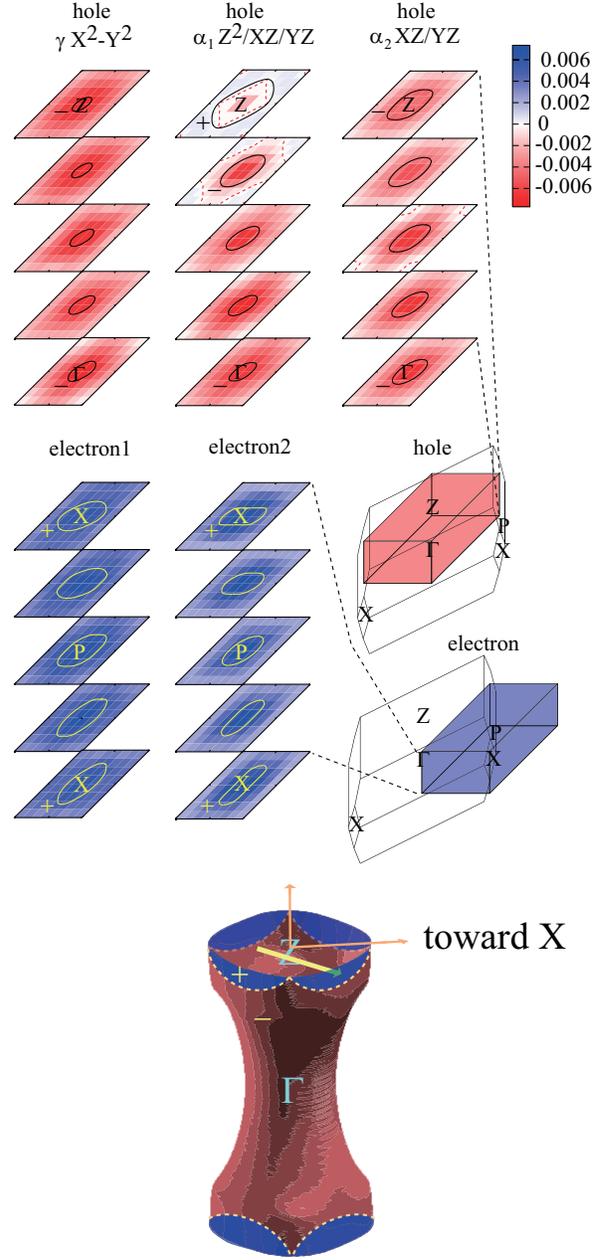}
\end{center}
\caption{(Color online) Contour plots of the gap function on the five Fermi surfaces
for five horizontal cuts around $\Gamma-$Z (hole) or X$-$P$-$X (electron).  
The solid lines represent the Fermi surfaces, while the dashed lines are the 
nodes of the gap.
The Brillouin zone is shown along with the regions where the 
contour plots of the gap are presented. In the bottom, a schematic view of the 
gap function on the $\alpha_1$=$Z^2/XZ/YZ$ hole Fermi surface is shown. The 
arrow bridging the Fermi surface 
indicates the intraband interaction that induces the sign change of the 
gap.
}
\label{fig4}
\end{figure}

We show the superconducting gap in Fig. \ref{fig4}.
The electron and hole Fermi surfaces have 
different signs of the gap for most portions, 
namely, the gap is basically $s\pm$, originating from the 
spin fluctuation mentioned above. 
The gap on the electron Fermi surfaces is fully open
since the  $X^2-Y^2$ $\gamma$ Fermi surface is present, 
as is the case for the 1111 systems with high pnictogen positions\cite{KKheight}.
On the other hand, as the volume of the $\alpha_1$ Fermi surface 
around the Z point grows upon lowering the pnictogen height by 
isovalent doping, a three-dimensional sign change of the gap 
takes place within this Fermi surface, as shown in Fig. \ref{fig4}.
The gap function on the $\alpha_1$ Fermi surface is schematically shown
at the bottom of Fig. \ref{fig4}. This sign change can be considered to be 
due to the repulsive intraband interaction 
within the $\alpha_1$ Fermi surface (shown by the arrow in the 
schematic figure), which becomes more effective as the 
volume of the Fermi surface around the Z point grows. 
In fact, we find that 
the plus-sign region of the gap on the $\alpha_1$ Fermi surface 
tends to shrink for smaller values of $x$.
A similar three-dimensional sign change 
of the superconducting gap has been found in a five orbital study of Ba122 in 
ref.\cite{Graser122} , although in the model of ref.\cite{Graser122} .
the $\alpha_1$ Fermi surface around the $Z$ point has an $X^2-Y^2$ character 
(or $xy$ in the notation of ref.\cite{Graser122} .) rather than $Z^2$. 

Since the three-dimensional nodes in the present study 
occur at portions of the Fermi surface having a strong $Z^2$ character,
the presence of the nodes does not strongly reduce $T_c$. 
This is in contrast to that in the case of LaFePO, where nodes of the 
gap, according to a similar RPA and other theoretical studies,
enters in the $X^2-Y^2$ and/or $XZ/YZ$ Fermi surfaces
\cite{KKheight,DHLee,Thomale,Ikeda1,Kariyado}, 
which play an important role in spin-fluctuation-mediated superconductivity.
The present view 
explains why $T_c$ is relatively high in BaFe$_2$(As$_{1-x}$P$_x$)$_2$ 
despite the presence of nodes in the superconducting gap. 

To summarize, we have obtained a three-dimensional ten-orbital model 
for the isovalent doping material BaFe$_2$(As$_{1-x}$P$_x$)$_2$, 
and applied RPA to obtain the 
superconducting gap originating from 
spin-fluctuation-mediated pairing. 
The obtained gap is basically $s\pm$, 
and we find no gap nodes 
on the electron Fermi surface because the 
$X^2-Y^2$ $\gamma$ Fermi surface persists up to a high P content.
On the other hand, there are three-dimensional nodes on the 
$Z^2/XZ/YZ$ portion of the hole Fermi surface. 
We presume that these three-dimensional 
nodes in the $Z^2/XZ/YZ$ hole Fermi surface are responsible for the 
experimentally observed nodal behavior of the superconducting gap of  
BaFe$_2$(As$_{1-x}$P$_x$)$_2$, since the presence of 
nodes in this material does not seem to strongly affect $T_c$. 
The relation between the present view and 
other experimental results will be of great interest.
A comparison between results of neutron scattering experiment and 
theoretical analysis based on the present gap structure 
is now in progress\cite{Ishikado}.
It is also necessary to understand how such a 
gap structure on the hole Fermi surface can be consistent with results of 
the ARPES experiment\cite{Shimojima}, 
or the apparently ``nodeless'' 
behavior observed in a specific 
heat experiment\cite{Kim}. We note here that our study 
does not completely rule out the possibility of nodes on the 
electron Fermi surface, since correlation effects that are not 
taken into account in the present ``band calculation+RPA'' study may 
give rise to gap nodes on the 
electron Fermi surface (even when the $X^2-Y^2$ hole Fermi surface 
is present)\cite{Ikeda1}.
Also, there is a possibility that in the {\it electron}-doped 
122 materials, where the $\gamma$ Fermi surface is less effective, 
the nodes may be on the electron Fermi surface, as suggested by 
a recent study\cite{Mazin122}.
In any case, the tendency of 
three-dimensional nodes entering the $Z^2/XZ/YZ$ hole Fermi surface of 
the 122 materials 
should remain as long as this Fermi surface is sufficiently large and 
the repulsive intraband interaction is effective.
In this context, a ``heavily hole-doped'' material, 
KFe$_2$As$_2$, mentioned as another nodal superconductor 
in the introductory part, is also of great interest. 
In fact, in our preliminary study of this material,  
we find a similar tendency of three-dimensional gap nodes 
entering the $Z^2/XZ/YZ$ hole Fermi surface.
In this study, we use a model that quantitatively 
reproduces the spin fluctuation modes observed 
in a recent neutron scattering experiment\cite{Lee}. 
The details will be published elsewhere.

We are grateful to 
Y. Matsuda, T. Shibauchi, S. Kasahara, K. Ishida, H. Ikeda, T. Shimojima, 
S. Shin, Y. Nagai, S. Shamoto, T. Yoshida, A. Fujimori, 
C. H. Lee, S. Onari, R. Arita, and H. Aoki for valuable discussions.
Numerical calculations were performed at the facilities of
the Information Technology Center, University of Tokyo, 
and also at the Supercomputer Center,
ISSP, University of Tokyo. 
This study has been partially supported by a
Grant-in-Aid for Scientific Research from  MEXT of Japan and from 
the Japan Society for the Promotion of Science.
%



\begin{thebibliography}{99}
\bibitem{review} For a review, see, e.g. K. Ishida, Y. Nakai and H. Hosono: 
J. Phys. Soc. Jpn. {\bf 78} (2009) 062001.
\bibitem{Matsuda} K. Hashimoto {\it et al.}: Phys. Rev. Lett. {\bf 102} 
(2009) 017002.
\bibitem{Ding} H. Ding {\it et al.}: Europhys. Lett. {\bf 83} (2008) 47001.
\bibitem{Fletcher} J. D. Fletcher {\it et al.}: 
Phys. Rev. Lett. {\bf 102} (2009) 147001.
\bibitem{Hicks} C. W. Hicks {\it et al.}: 
Phys. Rev. Lett. {\bf 103}  (2009) 127003.
\bibitem{Yamashita} M. Yamashita {\it et al.}: 
Phys. Rev. B {\bf 80} (2009) 220509(R).
\bibitem{LeeStruct} C. H. Lee {\it et al.}: J. Phys. Soc. Jpn. {\bf 77} 
(2008) 083704.
\bibitem{KKheight} K. Kuroki {\it et al.}: 
Phys. Rev. B {\bf 79} (2009) 224511.
\bibitem{KK1st}  K. Kuroki {\it et al.}: 
Phys. Rev. Lett. {\bf 101} (2008) 087004.
\bibitem{Mazin}  I. I. Mazin {\it et al.}: 
Phys. Rev. Lett. {\bf 101} (2008) 057003.
\bibitem{Ikeda0} H. Ikeda: J. Phys. Soc. Jpn. {\bf 77} (2008) 123707.
\bibitem{Nomura} T. Nomura: J. Phys. Soc. Jpn. {\bf 78} (2009) 034716.
\bibitem{Graser} S. Graser {\it et al.}: 
New J. Phys. {\bf 11} (2009) 025016.
\bibitem{Ikeda1}  H. Ikeda, R. Arita, and J. Kunes: 
Phys. Rev. B {\bf 81} (2010) 054502, also private communications regarding 
further studies.
\bibitem{DHLee} F. Wang, H. Zhai, and D.-H. Lee: Phys. Rev. B {\bf 81} (2010) 
184512.
\bibitem{Thomale} R. Thomale {\it et al.}: 
arXiv: 1002.3599.
\bibitem{Ren} S. Jiang {\it et al.}:
J. Phys.: Condens. Matter {\bf 21} (2009) 382203.
\bibitem{Kasahara} S. Kasahara {\it et al.}: 
Phys. Rev. B {\bf 81} (2010) 184519.
\bibitem{Hashimoto} K. Hashimoto {\it et al.}: 
Phys. Rev. B {\bf 81} (2010) 220501(R).
\bibitem{Nakai} Y. Nakai {\it et al.}: 
Phys. Rev. B. {\bf 81} (2010) 020503(R).
\bibitem{Shimojima} T. Shimojima: private communication.
\bibitem{Fukazawa} H. Fukazawa {\it et al.}: 
J. Phys. Soc. Jpn. {\bf 78} (2009) 083712.
\bibitem{Dong} J. K. Dong {\it et al.}: 
Phys. Rev. Lett. {\bf 104} (2010) 087005.
\bibitem{Furukawa} H. Kawano-Furukawa {\it et al.}: 
arXiv: 1005.4468.
\bibitem{Zhang} S. W. Zhang {\it et al.}: 
Phys. Rev. B {\bf 81} (2010) 012503.
\bibitem{Matsuda2} K. Hashimoto {\it et al.}: 
Phys. Rev. B {\bf 82} (2010) 014526. 
\bibitem{Sato} T. Sato {\it et al.}: 
Phys. Rev. Lett. {\bf 103} (2009) 047002.
\bibitem{Chubukov} A.V. Chubukov, M.G. Vavilov, and A.V. Vorontsov: 
Phys. Rev. B {\bf 80}  (2009) 140515(R).
\bibitem{Kontani} H. Kontani and S. Onari: Phys. Rev. Lett. {\bf 104} (2010) 
157001.
\bibitem{Ohno} Y. Yanagi {\it et al.}: 
Phys. Rev. B {\bf 82} (2010) 064518.
\bibitem{MaxLoc} N. Marzari and D. Vanderbilt: Phys. Rev. B 
{\bf 56} (1997) 12847; 
I. Souza, N. Marzari, and D. Vanderbilt: 
Phys. Rev. B {\bf 65} (2002) 035109.
The Wannier functions are generated by the code developed by
A. A. Mostofi, J. R. Yates, N. Marzari, I. Souza, and D. Vanderbilt,
(http://www.wannier.org/) 
for the energy window $-2.4$ eV $<\epsilon_k-E_F<$ 3.2eV,
where $\epsilon_k$ is the eigenenergy of the Bloch states
and $E_F$ the Fermi energy.
\bibitem{Mazin122} I. I. Mazin {\it et al.}: 
Phys. Rev. B {\bf 82} (2010) 180502.
\bibitem{Graser122} S. Graser, A. F. Kemper, T. A. Maier, H.-P. Cheng, 
P. J. Hirschfeld, and D. J. Scalapino: Phys. Rev. B {\bf 81} (2010) 214503.
\bibitem{Hirschfeld} P. J. Hirschfeld and D. J. Scalapino: Physics {\bf 3} 
(2010) 64.
\bibitem{pwscf} 
S. Baroni {\it et al.}:
http://www.pwscf.org/.
Here we adopt the exchange correlation functional introduced by
J. P. Perdew, K. Burke, and Y. Wang
(Phys. Rev. B {\bf 54} (1996) 16533), and the wave functions are expanded by 
plane waves up to a cutoff energy of 40 Ry.
8$^3$ $k$-point meshes are used.
\bibitem{struct} S. Kasahara: private communication.
\bibitem{Singh} D. J. Singh and M.-H. Du: Phys. Rev. Lett. {\bf 100} (2008) 
237003. 
\bibitem{Lebegue} S. Lebegue, Z. P. Yin, and W. E. Pickett: New J. Phys. 
{\bf 11} (2009) 025004.
\bibitem{GeorgesArita} V. Vildosola {\it et al.}: 
Phys. Rev. B {\bf 78} (2008) 064518.
\bibitem{Yoshida} T. Yoshida {\it et al.}: 
arXiv: 1008.2080.
\bibitem{Miyake} T. Miyake {\it et al.}: 
J. Phys. Soc. Jpn. {\bf 79} (2010) 044705.
\bibitem{Kariyado} T. Kariyado and M. Ogata: J. Phys. Soc. Jpn. {\bf 78} 
(2009) 043708.
\bibitem{Ishikado} M. Ishikado {\it et al.}: private communication.
\bibitem{Kim} J. S. Kim {\it et al.}: 
Phys. Rev. B {\bf 81} (2010) 214507.
\bibitem{Lee} C. H. Lee {\it et al.}:
arXiv: 1009.4001.
\end{thebibliography}
\end{document}